# Metric Structure and Dimensionality over a Borel Set via Uniform Spaces

*W.M. Stuckey*
*Dept of Physics*
*Elizabethtown College*
*Elizabethtown, PA  17022*
*stuckeym@etown.edu*

**Abstract**

We introduce a pregeometry that provides a metric and dimensionality over a Borel set (Wheeler's "bucket of dust") without assuming probability amplitudes for adjacency. Rather, a non-trivial metric is produced over a Borel set X per a uniformity base generated via the discrete topological group structures over X. We show that entourage multiplication in this uniformity base mirrors the underlying group structure. One may exploit this fact to create an entourage sequence of maximal length whence a fine metric structure. Unlike the statistical approaches of graph theory, this method can suggest dimensionality over low-order sets. An example over $Z_2 \times Z_4$ produces 3-dimensional polyhedra embedded in E4.

**Keywords:** Pregeometry, uniform spaces, graph theory

# 1. Introduction

According to Wheeler's pregeometry, properties of the spacetime manifold, such as metric, continuity, dimensionality, topology, locality, symmetry, and causality, might evolve mathematically from modeling underlying classical spacetime dynamics [1]. In an early attempt to derive dimensionality, Wheeler assigned probability amplitudes to the members of a Borel set[1] to stochastically establish adjacency [2]. He abandoned this idea, in part, because "too much geometric structure is presupposed to lead to a believable theory of geometric structure" [3]. In particular, he considered the manner in which probability amplitudes were assigned, and a metric introduced, to be *ad hoc*. However, recent models by Nagels [4], Antonsen [5], and Nowotny & Requardt [6] employing graph theory have, arguably, surmounted these objections.

In these approaches, assumptions concerning the stochastic analyses are minimal and natural. Nagels in reference 4, for example, assumes only that the order of the underlying set is large and the probability of adjacency is small. And, the notion of length per graph theory is virtually innate. Thus, Wheeler's pregeometric progeny have managed to obtain dimensionality, in various forms, over otherwise structureless sets.

Of course, one might contend that the connections between set members (bonds, links, edges) provide excessive *a priori* structure. However, connections are as fundamental to graph theory as the set itself. And we note, insofar as pregeometry is a search for a fundamental language of physics, the graph theoretical approach resonates with Heylighen's suggestion that a fundamental language of physics might be couched in "processes" [7]. His "events and arrows" parallel the "points and links" of the graph theoretical approaches *supra*. Antonsen in reference 5, for example, refers to his links as "interactions" and defines "corresponding 'second quantization' operators b, $b^+$." And, Requardt [8] refers to "dynamical bonds which transfer the elementary pieces of information among the nodes." It bodes well for the graph theoretical approach that a self-consistency relationship between spacetime geometry and physical processes – a pregeometric version of Einstein's equations – is already implied.

---

1. In reference 2, Wheeler describes the Borel set - "Loosely speaking, a Borel set is a collection of points which have not yet been assembled into a manifold of any particular dimensionality. Whether they are put together into lines, or surfaces, or volumes, or manifolds of higher dimensionality, or into odd combinations of such objects of varied dimensionality, is a matter of choice."

Given the promising nature of graph theory, we introduce a complementary approach. Whereby in graph theory the connection is fundamental to the metric and large-order sets are employed, our approach is valid over small-order sets and connections are inferred from the metric. This type of pregeometry may have bearing on the modeling of quantum non-locality where one might expect the structure of microscopic spacetime neighborhoods differs from that of macroscopic spacetime neighborhoods [9]. To contrast the statistical approach, we employ uniform spaces induced by discrete topological groups. In our approach, all subsets are open (discrete topology, as with graph theory [10]) and any group structure permissible. We begin by describing our construct of a uniformity base over a discrete topological group.

## 2. The Uniformity Base, Entourage Sequence and Metric

We refer to the underlying set X of order N as a 'Borel set', since it is denumerable, finite and structureless per Wheeler's description, and its members needn't be zero-dimensional objects. The category of uniform spaces contains topological groups [11] and a topological group may be created over any group with the discrete topology, so we assume all subsets of X are open and that we are free to consider any group structure G of order N.

We construct a uniformity base $U_B$ for the uniformity U via neighborhoods of the identity e of G per Geroch [12]. The entourage $A_\alpha$ of U is $\{(x, y) \in X \times X \mid xy^{-1} \in \alpha\}$ where $\alpha$ is a neighborhood of e in the topology over X. With X denumerable of order N, $\{(w, y) \in X \times X \mid w \neq y\}$ is partitioned equally into the entourages $A_x$ ($x \in X$ such that $x \neq e$) for the $N - 1$, order-two neighborhoods of e, i.e., $A_x$ is generated by $\{e, x\}$. The entourages $A_x$ and $\Delta \equiv \{(x, x) \mid x \in X\}$ constitute our base $U_B$ for U.

In order to produce a metric from this uniformity base, we borrow from a proof of the following theorem by Engelking [13].

"For every sequence $V_0, V_1, \ldots$ of members of a uniformity on a set S, where

$$V_0 = X \times X \text{ and } (V_{i+1})^3 \subset V_i \text{ for } i = 1, 2, \ldots,$$

there exists a pseudometric $\rho$ on the set S such that for every $i \geq 1$

$$\{(x, y) \mid \rho(x, y) < (½)^i\} \subset V_i \subset \{(x, y) \mid \rho(x, y) \leq (½)^i\}."$$

To find $\rho(x, y)$, consider all sequences of elements of S beginning with x and ending with y. For each adjacent pair $(x_n, x_{n+1})$ in any given sequence, find the smallest member of $\{V_i\}$ containing that pair. [The smallest $V_i$ will have the largest i, since $(V_{i+1})^3 \subset V_i$.] Suppose $V_m$ is that smallest member and let the 'artificial' distance between $x_n$ and $x_{n+1}$ be $(½)^m$. Summing for all adjacent pairs in a given sequence yields an 'artificial' distance between x and y for that particular sequence. According to the theorem, $\rho(x, y)$ is the greatest lowest bound obtained via the sequences.

Since our set X is denumerable and finite, this greatest lower bound will be non-zero and the result will be a metric. In order to obtain a metric with maximal resolution (fine metric) per this formalism, we need an entourage sequence $V_0, V_1, ...$ of maximal length. In order to produce such a sequence from the members of $U_B$, we note the following two properties.

First, not all the elements of our $U_B$ will be symmetric as required by Engelking's proof. In fact, $A_x$ is symmetric for all $x \in X$ such that $x = x^{-1}$. This, since for $(y, z) \in A_x$ such that $y \neq z$, $yz^{-1} = x$ and therefore, $zy^{-1} = x^{-1} = x \Rightarrow (z, y) \in A_x$. For the base members $A_x$ and $A_y$ such that $x = y^{-1}$, we have $A_x^{-1} = A_y$ where $A^{-1} = \{(w, z) \mid (z, w) \in A\}$. This, since for $(w, z) \in A_x$ such that $w \neq z$, $wz^{-1} = x$ and therefore, $zw^{-1} = x^{-1} = y \Rightarrow (z, w) \in A_y$. Accordingly, for $x = y^{-1}$ we must have $A_x \cup A_y$ appear in any $V_i$ to maintain the symmetry required by Engelking's theorem.

Second, we show that entourage multiplication of the $A_x \in U_B$ mirrors the underlying group structure. With $\Delta$ a subset of any entourage (uniquely and axiomatically), we have in general for entourages $A$ and $B$ that $A \subset AB$ and $B \subset AB$. Now consider $\{(x, y), (y, z) \mid (x, y) \in A_s$ and $(y, z) \in A_w$ with $x \neq y$ and $y \neq z\}$. In addition to $\Delta$, these account exhaustively for the elements of $A_s$ and $A_w$. For any such pair $(x, y)$ and $(y, z)$, $(x, z) \in A_s A_w$ by definition and $(x, z) \in A_{sw}$, since $sw = (xy^{-1})(yz^{-1}) = xz^{-1}$. The N pairs $(x, z)$ with $\Delta$ account exhaustively for the elements of $A_{sw}$ and, excepting the impact of $\Delta$ on $A_s A_w$, the N pairs $(x, z)$ account exhaustively for the elements of $A_s A_w$. Again, the impact of $\Delta$ on $A_s A_w$ is to render $A_s \subset A_s A_w$ and $A_w \subset A_s A_w$. Therefore, $A_s A_w = A_s \cup A_w \cup A_{sw}$. Accordingly, we may exploit subgroup structure to produce an entourage sequence of maximal length. Nested subgroup structures are particularly useful, since we require $(V_{i+1})^3 \subset V_i$. As an example of this approach, and to illustrate how it can suggest dimensionality, we obtain a fine metric over $Z_2 \times Z_4$.

## 3. Metric and Dimensionality over $Z_2 \times Z_4$

Let $X = \{a, b, c, e, A, B, C, D\}$ with the lower case elements those of the $Z_4$ subgroup where $b = b^{-1}$ and e is the identity. The upper case elements are $A = (1,a)$, $B = (1,b)$, $C = (1,c)$, and $D = (1,e)$. We have a nested subgroup structure of $Z_2 \subset Z_4 \subset Z_2 \times Z_4$, so we choose the following entourage sequence via $U_B$ :

$V_3 = A_b$

$V_2 = A_b \cup A_a \cup A_c$

$V_1 = A_b \cup A_a \cup A_c \cup A_D$

$V_0 = A_b \cup A_a \cup A_c \cup A_D \cup A_A \cup A_B \cup A_C$

which yields the following hierarchy for 'artificial' distances between pairs:

$\{(a,c), (b,e), (A,C), (B,D)\} \Rightarrow 1/8$ $\qquad [xy^{-1} = b]$

$\{(a,e), (b,a), (c,b), (e,c), (A,D), (B,A), (C,B), (D,C)\} \Rightarrow 1/4$ $\qquad [xy^{-1} = a \text{ or } c]$

$\{(a,A), (b,B), (c,C), (e,D)\} \Rightarrow 1/2$ $\qquad [xy^{-1} = D]$

$\{(a,D), (b,A), (c,B), (e,C), (A,e), (B,a), (C,b), (D,c)\} \Rightarrow 1$ $\qquad [xy^{-1} = B]$

$\{(a,C), (b,D), (c,A), (e,B), (A,c)\} \Rightarrow 1$ $\qquad [xy^{-1} = A \text{ or } C]$.

Therefore, we have the following metric structure (after a convenient renormalization):

| g(x,y) = 1 | g(x,y) = 2 | g(x,y) = 4 | g(x,y) = 5 | g(x,y) = 6 |
|---|---|---|---|---|
| (a,c) | (a,e) | (a,A) | (a,C) | (a,B) |
| (b,e) | (b,a) | (b,B) | (b,D) | (b,C) |
| (A,C) | (c,b) | (c,C) | (c,A) | (c,D) |
| (B,D) | (e,c) | (e,D) | (e,B) | (e,A) |
|  | (A,D) |  |  | (C,e) |
|  | (B,A) |  |  | (D,a) |
|  | (C,B) |  |  | (A,b) |
|  | (D,C) |  |  | (B,c) |

This is consistent with two, triangular polyhedra occupying E3's (figure 1) that are embedded in E4. In this embedding, the distances 1, 2, and 4 are understood as straight line (direct) distances while the distances 5 and 6 are obtained along indirect paths, e.g., g(e,A) = 6 obtains along the path e → a → A, since g(e,a) = 2 and g(a,A) = 4. [The path-dependent interpretation of g obtains with any embedding, tacitly if not explicitly.] Desideratum attained, we now speculate on a nexus to physics.

### 4. Nexus to Physics

Connections *a la* graph theory may be inferred from the edges of our embedded polyhedra and the direct paths between them. Not all of these connections need correspond to 'interactions/processes'. Rather, it may be that interactions obtain between events related by indirect metric distance, similar to the time-like exchange of particles. As with classical spacetime, the connections corresponding to direct paths might provide pure space-like or time-like relationships per the connections corresponding to interactions. In classical spacetime, for example, the structure of the space-like hypersurface constituting an observer's 'present' is inferred from the observer's past light cone. *Prima facie* this interpretation is deterministic, in contrast to quantum theory. However, stochasticity may obtain via alternative entourage sequences over X and/or deficient information concerning spatio-temporal boundary conditions.

For example, since $B = B^{-1}$ we could have chosen $V_1 = A_b \cup A_a \cup A_c \cup A_B$ in which case the columns $g(x,y) = 4$ and $g(x,y) = 5$ would have been switched in the metric *supra*. Heuristically, this ambiguity is reminiscent of the quantum state

$$|\psi\rangle = \frac{|\uparrow\rangle \pm |\downarrow\rangle}{\sqrt{2}}.$$

We might extend this speculation to the entangled state

$$|\psi\rangle = \frac{|\uparrow\downarrow\rangle \pm |\downarrow\uparrow\rangle}{\sqrt{2}}$$

via $Z_2 \times Z_2 \times Z_4$. In this case, we've the nested subgroup structure $Z_2 \subset Z_4 \subset Z_2 \times Z_4 \subset Z_2 \times Z_2 \times Z_4$, so we may use the entourage sequence

$V_5 = A_b$

$V_4 = A_b \cup A_a \cup A_c$

$V_3 = A_b \cup A_a \cup A_c \cup A_D$

$V_2 = A_b \cup A_a \cup A_c \cup A_D \cup A_A \cup A_B \cup A_C$

$V_1 = A_b \cup A_a \cup A_c \cup A_D \cup A_A \cup A_B \cup A_C \cup A_\delta$

$V_0 = X \times X$

where δ is the counterpart to D in the $Z_2$ duplication. Or, we may replace $V_1$ with $A_b \cup A_a \cup A_c \cup A_D \cup A_A \cup A_B \cup A_C \cup A_\beta$, where β is the counterpart to B in the $Z_2$ duplication. Either of these entourage sequences produces a fine metric whence two 'E3-polyhedra-embedded-in-E4' are embedded in E5. The choice of D or B in $V_3$ fixes *both* E4 substructures, so the connection between the two E4 subsets might be interrupted as a non-local, EPR-type[2] relationship in the otherwise global E4 spacetime structure of quantum mechanics (as opposed to an E5 embedding). However, the choice of δ or β in $V_1$ might be interpreted as fixing the temporal order of the measurements, which would allow the temporal ordering of space-like separated measurements to be ambiguous, as in the global M4 spacetime of special relativity. [In either case, we believe our formalism will – and any pregeometry must – address Monk's point [14] that quantum non-locality/non-separability occurs on macroscopic scales, so sub-Planck scale structures may not provide an ideal basis for pregeometry.]

In addition to originating with the ambiguity of entourage sequences, stochasticity may also surface via incomplete spatio-temporal boundary conditions. In any theory of physics, a spacetime region is described uniquely only after providing specifics concerning its spatio-temporal boundary. Our approach is truly spatio-temporal, i.e., it produces a 'block universe'[3]. Thus, we expect the unique description of a spacetime region will require information about its future boundary. An inability to supply this information may lead to stochastic, rather than deterministic, descriptions of the spacetime region(s) in question. In the entangled state *supra*, for example, knowing the temporal order of the measurements corresponds to specifying a particular spacetime foliation. This reduces ambiguity in the entourage sequence (fixes β or δ in $V_1$), thus reducing stochasticity. To determine the spatio-temporal description uniquely requires knowledge of a measurement outcome (fixes B or D in $V_3$). Thus, per this interpretation, the collapse of the quantum mechanical wave function is explicitly a two-step process (in contrast to quantum mechanics whereby the choice of a preferred spacetime foliation is tacit). And, wave function collapse is epistemological – not ontological – as one would expect in a 'block universe'.

---

2. EPR stands for Einstein-Podolsky-Rosen, who articulated problems stemming from quantum non-locality in a 1935 publication (*Phys. Rev.* **47**, 777).
3. A 'block universe' needn't be constructed dynamically, i.e., per the temporal evolution of a space-like hypersurface. However, it must subsume our dynamical perspective so as to be open to empirical investigation.

## 5. Conclusion

We have shown that metric structure and dimensionality may be induced over a Borel set per a uniformity base generated via discrete topological groups. The dimensionality produced herein is a simple embedding dimension, however it is inferred from the 'connections' between 'nodes'. Thus, should some of these connections ultimately correspond to physical interactions, our embedding dimension may be viewed in a dynamical spirit *a la* Requardt's 'connectivity-dimension' [15].

Our approach differs from graph theory in that connections are inferred from the metric, which obtains as a natural consequence of algebraic structure. Thus, a denumerable, finite set and all its possible group structures are the fundamental constituents of this formalism. The domain of discourse and group representations are unstipulated *a la* general covariance. This method is useful over low-order sets, which might be important in the pregeometric modeling of microscopic spacetime neighborhoods. In fact, there are intimations of both quantum statistics and M4 spacetime structure in this low-order approach. However, it may be difficult to employ with large-order sets, as there is no function that provides the number of group structures over a set of order N, unless N is prime. [In that case, $Z_N$ is the only available group structure.]

As with graph theoretical approaches, our model is 'bottom up' [16] in contrast to the 'top down' approach originally proposed by Sakharov [17] and developed more recently by Akama & Oda [18], and Terazawa [19]. In their models, particle physics is assumed fundamental to the spacetime geometry while we have assumed, per Wheeler, "the order of progress may not be physics $\rightarrow$ pregeometry, but pregeometry $\rightarrow$ physics" [20]. At this stage, it is probably best that researchers are engaged on all fronts.

**Figure 1**

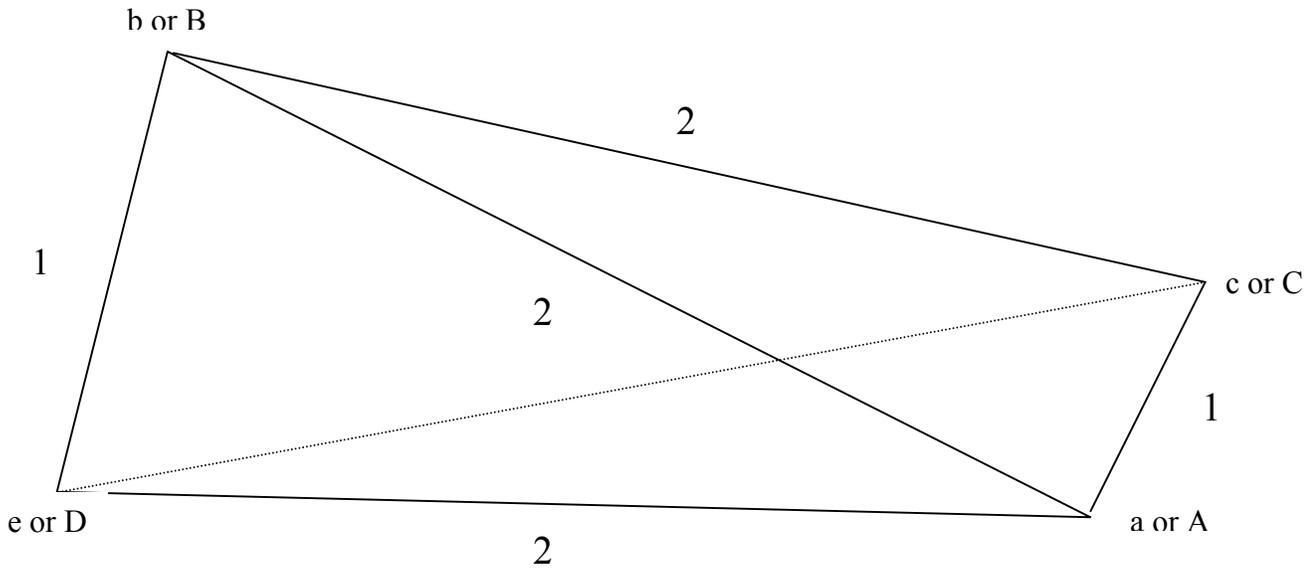